\documentclass{aa}
\usepackage{url}
\usepackage{graphicx}
\graphicspath{{./fig/}}

\newcommand{\EQ}{\begin{equation}}
\newcommand{\EN}{\end{equation}}
\newcommand{\EQA}{\begin{eqnarray}}
\newcommand{\ENA}{\end{eqnarray}}

\newcommand{\EEq}[1]{Equation~(\ref{#1})}
\newcommand{\Eq}[1]{Eq.~(\ref{#1})}
\newcommand{\Eqs}[2]{Eqs~(\ref{#1}) and~(\ref{#2})}

\newcommand{\Sec}[1]{Sect.~\ref{#1}}

\newcommand{\Fig}[1]{Fig.~\ref{#1}}
\newcommand{\FFig}[1]{Figure~\ref{#1}}

\newcommand{\bra}[1]{\langle #1\rangle}

\newcommand{\meanB}{\overline{B}}

\newcommand{\meanu}{\overline{u}}

\newcommand{\meanBB}{\overline{\vec{B}}}
\newcommand{\meanJJ}{\overline{\vec{J}}}

\newcommand{\meanuu}{\overline{\mbox{\boldmath $u$}}{}}{}
{}
{}
{}
{}
{}
{}
\newcommand{\uu}{{\vec{u}}}
\newcommand{\BB}{{\vec{B}}}
\newcommand{\JJ}{{\vec{J}}}

\newcommand{\ff}{\mbox{\boldmath $f$} {}}

\newcommand{\FF}{{\vec{F}}}

\newcommand{\kk}{{\vec{k}}}

\newcommand{\nab}{\vec{\nabla}}

\newcommand{\SSSS}{\mbox{\boldmath ${\sf S}$} {}}

\newcommand{\emf}{\mbox{\boldmath ${\cal E}$} {}}

\newcommand{\ii}{{\rm i}}

\newcommand{\DD}{{\rm D} {}}

\newcommand{\dd}{{\rm d} {}}

\def\la{\mathrel{\mathchoice {\vcenter{\offinterlineskip\halign{\hfil
$\displaystyle##$\hfil\cr<\cr\sim\cr}}}
{\vcenter{\offinterlineskip\halign{\hfil$\textstyle##$\hfil\cr<\cr\sim\cr}}}
{\vcenter{\offinterlineskip\halign{\hfil$\scriptstyle##$\hfil\cr<\cr\sim\cr}}}
{\vcenter{\offinterlineskip\halign{\hfil$\scriptscriptstyle##$\hfil\cr<\cr\sim\cr}}}}}
\def\ga{\mathrel{\mathchoice {\vcenter{\offinterlineskip\halign{\hfil
$\displaystyle##$\hfil\cr>\cr\sim\cr}}}
{\vcenter{\offinterlineskip\halign{\hfil$\textstyle##$\hfil\cr>\cr\sim\cr}}}
{\vcenter{\offinterlineskip\halign{\hfil$\scriptstyle##$\hfil\cr>\cr\sim\cr}}}
{\vcenter{\offinterlineskip\halign{\hfil$\scriptscriptstyle##$\hfil\cr>\cr\sim\cr}}}}}
\def\half{{\textstyle{1\over2}}}

\def\onethird{{\textstyle{1\over3}}}

\newcommand{\s}{\,{\rm s}}

\newcommand{\cm}{\,{\rm cm}}

\newcommand{\yjas}[3]{ #1, {J. Atmosph. Sci.,} {#2}, #3}

\newcommand{\ysph}[3]{ #1, {Sol. Phys.,} {#2}, #3}

\newcommand{\yan}[3]{ #1, {AN,} {#2}, #3}

\newcommand{\yana}[3]{ #1, {A\&A,} {#2}, #3}

\newcommand{\ygafd}[3]{ #1, {Geophys. Astrophys. Fluid Dyn.,} {#2}, #3}

\newcommand{\yphy}[3]{ #1, {Physica,} {#2}, #3}

\newcommand{\yprl}[3]{ #1, {Phys. Rev. Lett.,} {#2}, #3}

\newcommand{\ypr}[3]{ #1, {Phys. Rev.} {#2}, #3}

\newcommand{\yjcp}[3]{ #1, {J. Comput. Phys.,} {#2}, #3}
\newcommand{\yjour}[4]{ #1, {#2}, {#3}, #4}

\newcommand{\pproc}[4]{ #1, in {#2}, ed. #3 (#4) (in press)}

\begin{document}
\authorrunning{T. A. Yousef et al.}
\titlerunning{Turbulent magnetic Prandtl number}
\title{Turbulent magnetic Prandtl number and magnetic diffusivity
quenching from simulations}
\author{
T.\ A.\ Yousef\inst{1}, A.\ Brandenburg\inst{2} and G.\ R\"udiger\inst{3}}
\institute{
Department of Energy and Process Engineering,
Norwegian University of Science and Technology,
Kolbj{\o}rn Hejes vei 2B, N-7491 Trondheim, Norway
\and
NORDITA, Blegdamsvej 17, DK-2100 Copenhagen \O, Denmark
\and
Astrophysical Institute  Potsdam, An der Sternwarte 16, D-14482 Potsdam, Germany
}\date{Astron. Astrophys. 411, 321-327 (2003)\\
Recieved 21 February 2003 / Accepted 25 August 2003,~ $ $Revision: 1.51 $ $}

\abstract{
Forced turbulence simulations are used to determine the
turbulent kinematic viscosity, $\nu_{\rm t}$, from the
decay rate of a large scale velocity field.
Likewise, the turbulent magnetic diffusivity, $\eta_{\rm t}$,
is determined from the decay of a large scale magnetic field.
In the kinematic regime, when the field is weak, the turbulent
magnetic Prandtl number, $\nu_{\rm t}/\eta_{\rm t}$, is about unity.
When the field is nonhelical, $\eta_{\rm t}$ is quenched when
magnetic and kinetic energies become comparable.
For helical fields the quenching is stronger and can be described
by a dynamical quenching formula.
}

\maketitle
\section{Introduction}

The concept of turbulent diffusion is often invoked when modeling large
scale flows and magnetic fields in a turbulent medium.
Turbulent magnetic diffusion is similar to turbulent thermal diffusion
which characterizes the turbulent exchange of patches of warm and cold gas.
This concept is also applied to turbulent magnetic diffusion
which describes the turbulent exchange of patches of magnetic field
with different strengths and direction.
Reconnection of magnetic field lines is not explicitly required, but in the
long run unavoidable if the magnetic power spectrum is to decrease toward
small scales.
The idea of Prandtl is that only the energy carrying eddies contribute
to the mixing of large scale distributions of velocity and magnetic
field structures.
This leads to a turbulent magnetic diffusion coefficient
$\eta_{\rm t}\approx{1\over3}U\ell$,
where $U$ is the typical velocity and $\ell$ the
scale of the energy carrying eddies.
For the kinematic turbulent viscosity one expects similar values.
Analytic theory based on the quasilinear approximation
also produces similar (but not identical) values of $\eta_{\rm t}$
and $\nu_{\rm t}$ (e.g.\ Kitchatinov et al.\ 1994).

It is usually assumed that the values of $\eta_{\rm t}$
and $\nu_{\rm t}$ are independent of the molecular
(microscopic) viscosity and magnetic diffusivity, $\nu$ and $\eta$.
However, in the context of the geodynamo or in laboratory liquid
metals the microscopic magnetic Prandtl number,
$P_{\rm m}=\nu/\eta$ is very small ($\approx10^{-5}$).
This has dramatic consequences for the magnetorotational
instability (Balbus \& Hawley 1991).
This instability is generally accepted as the main mechanism producing
turbulence in accretion discs (Balbus \& Hawley 1998).
For sufficiently small values of $P_{\rm m}$, however,
this instability is suppressed (R\"udiger \& Shalybkov 2002).
On the other hand, the Reynolds number of the flow is quite large
($10^5\ldots10^6$) and the flow therefore most certainly turbulent.
This led Noguchi et al.\ (2002) to invoke a turbulent kinematic
viscosity, $\nu_{\rm t}$, but to retain the microscopic value of $\eta$.
The resulting {\it effective} magnetic Prandtl number they used was $10^{-2}$
-- big enough for the magnetorotational instability to develop.
On may wonder, of course, why one should not instead use
turbulent values for both coefficients, i.e.\
$\nu_{\rm t}/\eta_{\rm t}\approx1$. This would lead to even more
favorable conditions for the magnetorotational instability
(R\"udiger et al.\ 2002).

Similar constraints have also been reported for the
convection-driven geodynamo: Christensen et al.\ (1999) found
that there is a minimum value of $P_{\rm m}$ of about 0.25 below
which dynamo action does not occur at all.
Similar results have also been reported by Cattaneo (2003).
These results are disturbing, because both for the sun and for the
earth, $P_{\rm m}\ll1$.
For $P_{\rm m}$ of order unity, on the other hand,
earth-like magnetic configurations can more easily be reproduced
(see Kutzner \& Christensen 2002).

Because of these restrictions, one wonders whether the effective
magnetic Prandtl number to be used is not $P_{\rm m}$, but rather
the turbulent value, $P_{\rm m,t}=\nu_{\rm t}/\eta_{\rm t}$.
This raises the important questions whether $P_{\rm m,t}$ is actually
of order unity and whether it is independent of the microscopic value,
$P_{\rm m}$.
The aim of this paper is to estimate the value of $P_{\rm m,t}$
using turbulence simulations.

The knowledge of
the value of $P_{\rm m,t}$ is also important for the solar dynamo.
The qualitative properties of the dynamo depend on the relative
importance of the large scale flows and hence on the
magnitude of $\eta_{\rm t}$. If $\eta_{\rm t}$ is too large,
the influence of a meridional flow of say 10 m/s is small so
that only little modification can be expected for the basic
$\alpha\Omega$-dynamo (Roberts \& Stix 1972).
In this case, however, we know that conventional
dynamo models of the solar activity cycle have difficulty to explain
Sp\"orer's law of equatorward sunspot migration. The alternative that the
resulting poleward migration can be overcompensated by an internal
equatorward flow requires a sufficiently small value of $\eta_{\rm t}$,
which implies that $P_{\rm m,t}>1$ (Choudhuri et al.\ 1995; Dikpati \& 
Charbonneau 1999; Bonanno et al.\ 2002).

Given the importance of the value of the turbulent magnetic Prandtl
number it is useful to assess the problem using three-dimensional simulations
of turbulent flows.
We determine $\nu_{\rm t}$ and $\eta_{\rm t}$ by measuring the decay
rate of a large scale (mean) velocity and magnetic field, $\meanuu$ and
$\meanBB$, respectively.
We emphasize that we are not addressing the question whether
$\nu_{\rm t}$ and $\eta_{\rm t}$ can really be used in studies
of the dynamo or the magnetorotational instability, for example.

We consider weakly compressible nonhelically forced turbulence and
use a model similar to that of Brandenburg (2001), but with kinetic
helicity fluctuating about zero.
Dynamo action for such a model has recently been considered by
Haugen et al.\ (2003), but it sets in only at magnetic Reynolds
numbers above $\sim30$, which is not the case in the present simulations.
We begin however by first reviewing the basic results for the
values of $\nu_{\rm t}$ and $\eta_{\rm t}$ within the framework
of the quasilinear (Roberts \& Soward 1975, R\"udiger 1989) and
other approximations.

\section{Results from quasilinear approximation}

For steady homogeneous isotropic turbulence the correlation tensor
is independent of $\vec{x}$ and $t$, i.e.\
\begin{equation}
\langle u_i'(\vec{x},t)u_j'(\vec{x}+\vec{\xi},t+\tau)\rangle
=Q_{ij}(\vec{\xi},\tau),
\end{equation}
where angular brackets denote an ensemble average and
primes fluctuations about the average.
In the quasilinear approximation the transport coefficients are
conveniently expressed in terms of the Fourier transformed correlation
tensor, $\hat Q_{ij}(\vec{k},\omega)$, which is normalized such that
\begin{equation}
Q_{ij}(\vec{\xi},\tau)=\int\!\!\int\hat Q_{ij}(\vec{k},\omega)
e^{i(\vec{k}\cdot\vec{\xi}-\omega\tau)}d\vec{k}\,d\omega.
\end{equation}
For the turbulent viscosity and the turbulent magnetic diffusivity
one finds respectively (R\"udiger 1989)
\begin{equation}
\nu_{\rm t} = {4\over 15} \int\!\!\int {\nu^3 k^6 \hat Q_{ll}(\vec{k},\omega)\over
(\omega^2 + \nu^2 k^4)^2} d\vec{k}\,d\omega,
\label{nuet}
\end{equation}
\begin{equation}
\eta_{\rm t} = {1\over 3} \int\!\!\int {\eta k^2 \hat Q_{ll} (\vec{k},
\omega) \over \omega^2 + \eta^2 k^4} d\vec{k}\,d\omega.
\label{etat}
\end{equation}
Obviously, both quantities
are of the same order of magnitude, but they are not identical. In the limits $\nu,\eta
\to 0$ the expressions are drastically simplified, i.e.
\begin{equation}
\nu_{\rm t} = {1\over 15} \int\limits_{-\infty}^\infty \langle\vec{u}'(\vec{x},t)
\cdot\vec{u}'(\vec{x},t+\tau)\rangle d\tau
\label{nuet1}
\end{equation}
and
\begin{equation}
\eta_{\rm t} = {1\over 6} \int\limits_{-\infty}^\infty \langle\vec{u}'(\vec{x},t)
\cdot\vec{u}'(\vec{x},t+\tau)\rangle d\tau ,
\label{etat1}
\end{equation}
so that for the turbulent magnetic Prandtl number is
\begin{equation}
P_{\rm m,t} = {\nu_{\rm t} \over \eta_{\rm t}} = {2\over 5}=0.4.
\label{Pm}
\end{equation}
This results is similar to that of Nakano et al.\ (1979) for the
thermal Prandtl number.

R\"udiger (1989) lists a number of other approaches for calculating
turbulent transport coefficients, which all yield Prandtl numbers around
or below unity.
One particular approach
is the renormalization group analysis which was applied to turbulence
by Forster et al.\ (1977) for the case of a passive scalar, and later
by Fournier et al.\ (1982) to the case with magnetic fields.
These results are valid in the long-time large-scale limit, and the value
of $P_{\rm m,t}$ turned out to be close to 0.7; see Eq.~(23) of
Fournier et al.\ (1982).

Kitchatinov et al.\ (1994) use a mixing length approximation where
terms of the form $\dd/\dd t-\nu\nabla^2$ are replaced by
$\tau_{\rm corr}^{-1}$, where $\tau_{\rm corr}$ is the correlation
time of the turbulence.
They find $\nu_{\rm t}=(4/15)\tau_{\rm corr}u_{\rm rms}^2$ and
$\eta_{\rm t}=(1/3)\tau_{\rm corr}u_{\rm rms}^2$, so
$P_{\rm m,t}=4/5=0.8$.
Yet another approach is the $\tau$-approximation where triple
correlations are replaced by a damping term that is proportional to the
quadratic moments (e.g.\ Kleeorin et al.\ 1996, Blackman \& Field 2002).
Here no Fourier transformation in time is used.
This gives, as before, $\eta_{\rm t}=(1/3)\tau u_{\rm rms}^2$
(where $\tau$ is now interpreted as a relaxation time), but
$\nu_{\rm t}=(2/15)\tau u_{\rm rms}^2$, so $P_{\rm m,t}=2/5=0.4$.
This is half the value obtained from the mixing length approximation,
but the same as in \Eq{Pm}.

The fact that in all these cases $P_{\rm m,t}$ is less than unity can be
traced back to the presence of the pressure term in the momentum equation.
If this term is ignored (as in pressureless Burgers turbulence or
`burgulence') one always gets $P_{\rm m,t}=1$.

It is tempting to speculate that the discrepancy between the
different analytic approaches is related to the validity of some
idealizing assumptions made in order to apply the quasilinear
and other approximations.
Clearly, additional approaches are needed to get a more complete picture
regarding the correct value of $P_{\rm m,t}$.
It is nevertheless encouraging that $P_{\rm m,t}$ does not strongly
deviate from unity.

In the remainder of this paper we estimate $\nu_{\rm t}$ and
$\eta_{\rm t}$ numerically by considering the decay of an initial large
scale velocity or magnetic field, respectively, in the presence of small
scale turbulence.

\section{The model}

The equations describing compressible isothermal hydromagnetic flows
with constant sound speed, $c_{\rm s}$, are
\EQ
{\DD\uu\over\DD t}=-c_{\rm s}^2\nab\ln\rho+{\JJ\times\BB\over\rho}
+\FF_{\rm visc}+\ff,
\label{dudt}
\EN
\EQ
{\DD\ln\rho\over\DD t}=-\nab\cdot\uu,
\EN
\EQ
{\partial\BB\over\partial t}=\nab\times(\uu\times\BB)+\eta\nabla^2\BB,
\label{dAdt}
\EN
where $\uu$ is the velocity, $\rho$ the density, $\BB$ is the magnetic field,
and $\JJ=\nab\times\BB/\mu_0$ is the current density with $\mu_0$ being
the vacuum permeability. The viscous force is
\EQ
\FF_{\rm visc}=\nu\left(\nabla^2\uu+\onethird\nab\nab\cdot\uu
+2\SSSS\cdot\nab\ln\rho\right),
\EN
where ${\sf S}_{ij}=\half(u_{i,j}+u_{j,i})-\onethird\delta_{ij}
\nab\cdot\uu$ is the traceless rate of strain tensor.

We solve the equations using the Pencil Code\footnote{
\url{http://www.nordita.dk/data/brandenb/pencil-code}}, which is a
memory-efficient sixth-order finite difference code using the
$2N$-RK3 scheme of Williamson (1980).
For most of the simulations a resolution of $128^3$ meshpoints is
used, but in \Sec{Sindependence} a higher resolution of up to $512^3$
meshpoints was necessary.

We focus on the case where the forcing, $\ff$, occurs at a wavenumber
around $k_{\rm f}=10$.
The forcing is such that the turbulence is subsonic and nonhelical.
We consider two different periodic initial conditions,
\EQ
\BB=(\cos k_1z,0,0)B_0
\quad\mbox{(nonhelical)}
\EN
and
\EQ
\BB=(\cos k_1z,\sin k_1z,0)B_0
\quad\mbox{(helical)},
\EN
where $B_0$ is the amplitude of the initial field.
In the fully helical case one may
expect a different decay time because the magnetic helicity is a
conserved quantity in the limit of small magnetic diffusivity.
For the velocity field we use similar initial conditions, but we
do not expect this to be sensitive to helicity, because kinetic
helicity is not conserved in the limit $\nu\to0$, and would only
be conserved in the unphysical case $\nu=0$.

A detailed discussion of the initial conditions may at first glance appear
somewhat surprising, because for forced turbulent flows the
initial conditions are normally forgotten after about one turnover time.
This is indeed the case for hydrodynamic turbulence, but not for
hydromagnetic turbulence if the magnetic field has net magnetic helicity.
The reason is that, regardless of the level of turbulence, the net
magnetic helicity can only change on the resistive time scale.
Our results below confirm this and they are indeed in agreement
with earlier model predictions (cf.\ Blackman \& Brandenburg 2002).
The situation would be different if the initial field was bi-helical,
i.e.\ with oppositely helical contributions at different scales.
This case has been studied elsewhere (Yousef \& Brandenburg 2003).

\begin{figure}[t!]\centering\includegraphics[width=0.45\textwidth]{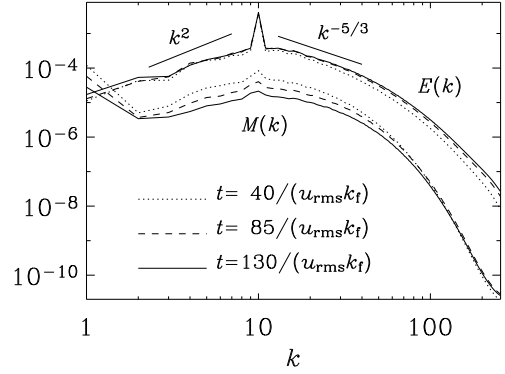}\caption{
Kinetic and magnetic energy spectra at three different times
for a run with $\mbox{Re}=150$ and $R_{\rm m}=15$.
}\label{Fetat512a}\end{figure}

In \Fig{Fetat512a} we show kinetic and magnetic energy spectra of the
run with $\mbox{Re}=150$ and $R_{\rm m}=15$ at three different times
using a resolution of $512^3$ meshpoints.
The kinetic energy shows indications of a short inertial range in
$15<k<40$.
Below the forcing scale, in $2<k<9$, velocity and magnetic fields are
random and $\delta$-correlated in space, giving rise to a $k^2$ spectrum.
The magnetic energy is substantially weaker than the kinetic energy.
This is because here the magnetic Prandtl number is small, $P_{\rm m}=0.1$,
and the magnetic Reynolds number is subcritical for dynamo action.
With our definition of $R_{\rm m}$ the critical value lies around 25
(Haugen et al.\ 2003).
The small scale magnetic energy is therefore maintained by constantly
stirring the slowly decaying large scale field.

Given that the initial large scale field depends only on $z$, it makes
sense to define a mean field by averaging over the $x$ and $y$ directions.
Alternatively, one might define an average by Fourier filtering, but
this has the disadvantage that not all the Reynolds rules are satisfied.
For example, the average of a product of a mean and a fluctuating quantity
would not vanish.
However, for all practical purposes our horizontal average is nearly
equivalent to a projection onto the $k=k_1$ Fourier mode. 
Indeed, the main reason for forcing at a large wavenumber, $k_{\rm f}=10$,
is that we need some degree of scale separation.
Without scale separation, there would be no way of distinguishing
between mean and fluctuating fields.
Since the velocity fluctuations are constantly driven via the
forcing term, it would be
impossible to measure any decay of the mean velocity.
Nevertheless, even with scale separation there will always be
a certain level of noise in the mean field whose energy is
$(k_1/k_{\rm f})^2$ times smaller than energy of the fluctuations.
This means that we can measure an exponential decay of the mean field only in a
certain window where nonlinear effects are already weak, but were
the noise level is not yet reached.

\section{Results}

\subsection{Decay of $\meanu$ and $\meanB$}

\begin{figure}[t!]\centering\includegraphics[width=0.45\textwidth]{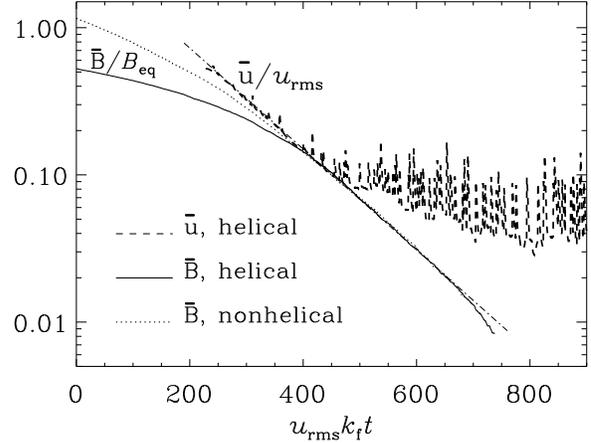}\caption{
Decay of large scale helical velocity and magnetic fields
(dashed and solid lines, respectively).
The graph of $\meanuu(t)$ has been shifted so that both $\meanuu(t)$
and $\meanB(t)$ share the same tangent (dash-dotted line), whose
slope corresponds to
$\nu_{\rm t}=\eta_{\rm t}=0.86u_{\rm rms}/k_{\rm f}$.
The decay of a nonhelical magnetic field is shown for
comparison (dotted line).
}\label{Fpn_comp_prm}\end{figure}

We begin by considering the decay of a helical large scale magnetic field
and compare it with the decay of a large scale helical velocity field
in a purely hydrodynamic simulation; see \Fig{Fpn_comp_prm}.
Here, large scale velocity and magnetic fields are defined as horizontal
averages over $x$ and $y$; the result is denoted by $\meanuu$ and
$\meanBB$, respectively.
During the time interval when mean velocity and magnetic field decay
exponentially, the corresponding decay rates are determined as
\EQ
\lambda_u(\meanuu)={\dd\ln\bra{\meanuu^2}^{1/2}\over\dd t},\quad
\lambda_B(\meanBB)={\dd\ln\bra{\meanBB^2}^{1/2}\over\dd t}.
\EN
In the graphs of $\lambda_u(\meanu)$ and $\lambda_B(\meanB)$ an
exponential decay shows up as a plateau.
The magnetic field decay is initially slow, so $\lambda_B(\meanB)$ is
initially not constant, but then it speeds up and $\lambda_B(\meanB)$
reaches a plateau.
The decay of the velocity field is immediately fast
and $\lambda_u(\meanu)$ lies immediately on a plateau.
This suggests that the turbulent magnetic diffusivity is affected
by the strong initial field that in turn gives rise to a quenching of the
turbulent magnetic diffusivity.
Strong means that the magnetic field strength is comparable with the
equipartition field strength, $B_{\rm eq}=\bra{\mu_0\rho\uu^2}^{1/2}$.
The initially strong large scale flow and the associated vorticity,
on the other hand, do not and are also not expected to affect the
turbulent viscosity and the associated decay of this large scale flow.
For $|\meanBB|\ll B_{\rm eq}$, however, both $\meanuu$ and $\meanBB$
decay at the same rates, $\lambda_u$ and $\lambda_B$, respectively.
This allows us to calculate
\EQ
\nu_{\rm t}=\lambda_u/k_1^2,\quad
\eta_{\rm t}=\lambda_B/k_1^2,
\EN
where $k_1$ is the wavenumber of the initial large scale velocity
and magnetic fields.
 From the present simulations, where $k_{\rm f}/k_1=10$, we find
\EQ
\nu_{\rm t}\approx\eta_{\rm t}=(0.8\ldots0.9)\times u_{\rm rms}/k_{\rm f}
\quad\mbox{(for $\meanBB^2\ll B_{\rm eq}^2$)}.
\label{nut_etat}
\EN
Once $|\meanuu|$ has decreased below a certain level ($<0.1u_{\rm rms}$),
it cannot decay further and continues to fluctuate around $0.08u_{\rm rms}$,
corresponding to the level of the rms velocity of the (forced!) turbulence at $k=k_1$
(see the dashed line in \Fig{Fpn_comp_prm}).

\begin{figure}[t!]\centering\includegraphics[width=0.45\textwidth]{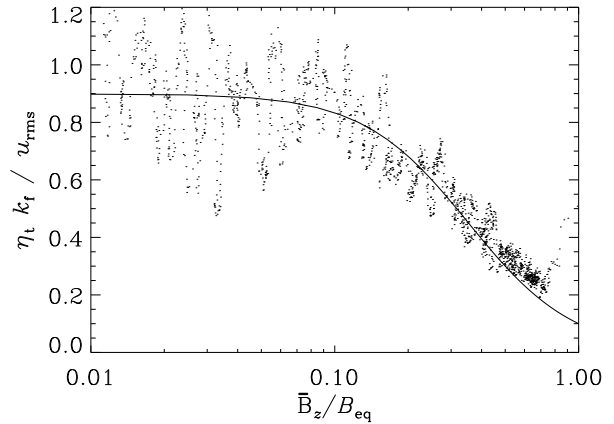}\caption{
Dependence of the turbulent diffusion coefficient on the magnitude of
the mean field. The initial field is helical and corresponds to data
points on the right hand side of the plot. $R_{\rm m}\approx20$.
The data are best fitted by $a=8=0.4R_{\rm m}$.
}\label{Fpdecay_law}\end{figure}

The quenching of the magnetic diffusivity,
$\eta_{\rm t}=\eta_{\rm t}(\meanBB)$, can be obtained
from one and the same run by simply determining the decay rate,
$\lambda_B(\meanB)$, at different times, corresponding
to different values of $\meanB=|\meanBB|$; see \Fig{Fpdecay_law}.
To describe departures from purely exponential decay we adopt a
$\meanBB$-dependent $\eta_{\rm t}$ expression of the form
\EQ
\eta_{\rm t}(\meanBB)=\eta_{\rm t0}/(1+a\meanBB^2/B_{\rm eq}^2),
\label{etatB}
\EN
where $\eta_{\rm t0}$ is the unquenched (kinematic) value of
$\eta_{\rm t}$, described approximately by \Eq{nut_etat}, and
$a$ is a fit parameter.
According to Cattaneo \& Vainshtein (1991) the parameter $a$
is expected to be of the order of the
magnetic Reynolds number based on the microscopic magnetic diffusivity,
\EQ
R_{\rm m}=u_{\rm rms}k_{\rm f}/\eta.
\label{Rm}
\EN
\FFig{Fpdecay_law} suggests that $a\approx0.4R_{\rm m}$.

\begin{figure}[t!]\centering\includegraphics[width=0.45\textwidth]{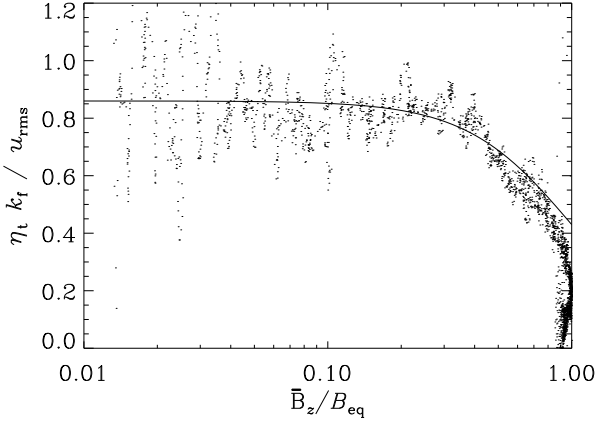}\caption{
Dependence of the turbulent diffusion coefficient on the magnitude of
the mean field. The initial field is nonhelical. $R_{\rm m}\approx20$.
The data are best fitted by $a=1$, independent of $R_{\rm m}$.
}\label{Fetat128e}\end{figure}

Before we discuss the effective quenching behavior of $\eta_{\rm t}$
in more detail we should note that \Eq{etatB}, and in particular the
value of $a$, do not apply universally and depend on the field geometry.
This is easily demonstrated by considering a nonhelical initial field.
In that case the decay becomes unquenched already for
$\meanBB^2/B_{\rm eq}^2\approx1$.
\EEq{etatB} can still be used as a reasonable fit formula, but now $a=1$
produces a good fit (independent of $R_m$); see \Fig{Fetat128e}.

In the nonhelical case there is an initial phase where the field
increases due to the wind-up of the large scale field.
Since we measure $\eta_{\rm t}$ from the decay rate of the large scale
field, this would formally imply negative values of $\eta_{\rm t}$.
Traces of this effect can still be seen in \Fig{Fetat128e}
near $\meanBB^2/B_{\rm eq}^2=1$.
For this reason our method can
only give reliable results if $|\meanBB|\la 0.8B_{\rm eq}$.
In the case of a helical initial field, on the other hand, we have
$\meanJJ\times\meanBB=0$, i.e.\ the large scale field is force-free
and interacts only weakly with the turbulence.
In particular, there is no significant amplification from
the initial wind-up of the large scale field.

\subsection{Comparison with the dynamical quenching model}
\label{Sdynquench}

In the case of a helical field and for
$\meanBB^2/B_{\rm eq}^2\ga R_{\rm m}^{-1}$
the slow decay of $\meanBB$
is related to the conservation of magnetic helicity.
As discussed already by Blackman \& Brandenburg (2002),
this behavior is related to the phenomenon of selective
decay (e.g.\ Montgomery et al.\ 1978) and can be described by the
dynamical quenching model.
This model goes back to an early paper by Kleeorin \& Ruzmaikin (1982,
see also Kleeorin et al.\ 1995),
but it applies even to the case where the turbulence is nonhelical and
where there is no $\alpha$ effect in the usual sense.
However, the magnetic contribution to $\alpha$ is still
non-vanishing because it is
driven by the helicity of the large scale field.

To demonstrate this quantitatively we solve,
in the one mode approximation ($\kk=\kk_1$) with
$\meanBB=\hat\BB\exp(\ii\kk_1z)$, the mean-field induction equation
\EQ
{\dd\hat\BB\over\dd t}=\ii\kk_1\times\hat{\emf}
-\eta k_1^2\hat\BB
\label{ODE1}
\EN
together with the dynamical $\alpha$-quenching formula
[Eq.~(13) of Blackman \& Brandenburg (2002)]
\EQ
{\dd\alpha\over\dd t}=-2\eta k_{\rm f}^2
\left(\alpha+\tilde{R}_{\rm m}{{\rm Re}(\hat{\emf}^*\cdot\hat\BB)
\over B_{\rm eq}^2}\right),
\label{ODE2}
\EN
where
\EQ
\hat{\emf}=\alpha\hat\BB-\eta_{\rm t}\ii\kk_1\times\hat\BB
\label{hat_emf}
\EN
is the
electromotive force, and $\tilde{R}_{\rm m}$ is defined as the ratio
$\eta_{\rm t0}/\eta$, which is expected to be close to the value of
$R_{\rm m}$ as defined by \Eq{Rm}.

\begin{figure}[t!]\centering\includegraphics[width=0.45\textwidth]{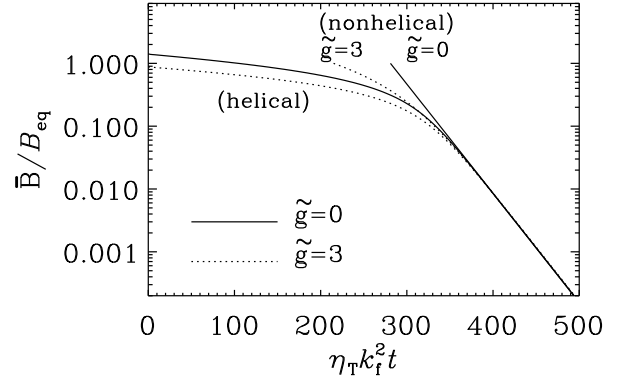}\caption{
Dynamical quenching model with helical and nonhelical initial fields.
The quenching parameters are $\tilde{g}=0$ (solid line) and 3 (dotted line).
The graph for the nonhelical cases has been shifted in $t$ so that one
sees that the decay rates are asymptotically equal at late times.
}\label{Fpselected_decay}\end{figure}

In \Fig{Fpselected_decay} we show the evolution of $\meanBB/B_{\rm eq}$
for helical and nonhelical initial conditions, $\hat\BB\propto(1,\ii,0)$
and $\hat\BB\propto(1,0,0)$, respectively.
In the case of a nonhelical field, the decay rate is not quenched at all,
but in the helical case quenching sets in for
$\meanBB^2/B_{\rm eq}^2\ga R_{\rm m}^{-1}$.

In the helical case, the onset of quenching at
$\meanBB^2/B_{\rm eq}^2\approx R_{\rm m}^{-1}$ is well reproduced by
the simulations.
In the nonhelical case, however, some weaker form of quenching sets in
when $\meanBB^2/B_{\rm eq}^2\approx1$ (\Fig{Fetat128e}).
We refer to this as standard quenching (e.g.\ Kitchatinov et al.\ 1994)
which is known to be always present.
In Blackman \& Brandenburg (2002) this was modeled by allowing
in \Eq{hat_emf} $\eta_{\rm t}$ to be $\meanBB$-dependent.
They adopted the formula
\EQ
\eta_{\rm t}=\eta_{\rm t0}/(1+\tilde{g}|\bra{\meanBB}|/B_{\rm eq})
\EN
and found that, for a range of different values of $R_{\rm m}$,
$\tilde{g}=3$ resulted in a good description of
the simulations of cyclic $\alpha\Omega$-type dynamos
(Brandenburg et al.\ 2002).
We emphasize that this $\eta_{\rm t}$ is {\it not} used in a diagnostic
way as in \Eq{etatB}, but rather in the numerical solution of
\Eqs{ODE1}{ODE2}.
The resulting decay law, shown as a dotted line in
\Fig{Fpselected_decay}, agrees now with the decay law seen in the
turbulence simulations (\Fig{Fpn_comp_prm}).
The helical case with $\tilde{g}=3$ is still compatible with the
simulations.

\section{Independence of microscopic viscosity}
\label{Sindependence}

\begin{figure}[t!]\centering\includegraphics[width=0.45\textwidth]{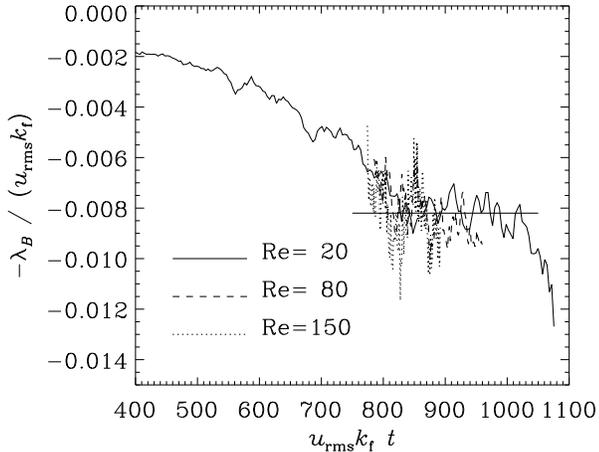}\caption{
Decay rate for three different values of $\mbox{Re}$ and $R_{\rm m}=20$
(fixed), corresponding to values of $P_{\rm m}=R_{\rm m}/\mbox{Re}$
ranging from 0.1 to 1.
All three curves have a plateau where the value of $\lambda_{\rm B}$
is the same.
For $R_{\rm m}=80$ and 150 the graphs of $\lambda_B$ have been shifted in
$t$ so that all three graphs show the plateau in approximately the same
time interval.
}\label{Fpn_comp_prm2_der}\end{figure}

Finally we need to show that the turbulent magnetic Prandtl number
is indeed independent of the microscopic magnetic Prandtl number.
In \Fig{Fpn_comp_prm2_der} we plot the decay rates, obtained by
differentiating $\ln\meanBB(t)$, for three different
values of the microscopic viscosity, keeping $\eta$ fixed.
The resulting values of the flow Reynolds number,
$\mbox{Re}=u_{\rm rms}k_{\rm f}/\nu$, vary between 20 and 150,
giving $P_{\rm m}$ in the range between 0.1 and 1.
Within plot accuracy the three values of $\lambda_B$ turn out
to be identical in the interval where the decay is exponential.

The duration of this interval is
$u_{\rm rms}k_{\rm f}\Delta t\approx200$, which is comparable
to the time interval in \Fig{Fpn_comp_prm} during which the
decay is exponential.
In one of the three cases ($R_{\rm m}=20$) the initial field
was rather strong, so that it takes a long time before the
magnetic helicity constraint becomes unimportant so that
the field can decay exponentially
($u_{\rm rms}k_{\rm f}t\approx800$).

The numerical resolution used in most of the models is $128^3$ mesh points.
However, as Re is increased, higher resolution is required.
For Re=80 we used $256^3$ mesh points and
for Re=150 we used $512^3$ mesh points.
This implies mesh a Reynolds number, $u_{\rm rms}\Delta x/\nu$,
based on the mesh spacing $\Delta x$, of about 18.
Empirically we know that larger values are not generally possible.

\section{Conclusions}

The turbulence simulations presented here have shown that the turbulent
magnetic Prandtl number is always of order unity, regardless of the
values of the {\it microscopic} magnetic Prandtl number.
Under the assumption of incompressibility, both the quasilinear
approximation and the renormalization group approach give turbulent
magnetic Prandtl numbers somewhat below unity, which is related to the
pressure term in the momentum equation.
Here we find instead $P_{\rm m,t}\approx1$.
There are several plausible reasons for this discrepancy:
(i) our simulations are actually weakly compressible,
(ii) they are non-steady and,
(iii) the idealizing assumptions made in the analytic approaches
may not be justified.

Our results have also shown that, for nonhelical magnetic fields,
the turbulent magnetic diffusivity is quenched when the magnetic energy
becomes comparable to the kinetic energy.
For helical magnetic fields, however, an apparent suppression of the
decay rate is observed which agrees with predictions from a dynamical
quenching model.
If this suppression is described by an algebraic expression, quenching
would set in for magnetic energies much below the kinetic energy.

The present work demonstrates that the dynamical quenching approach
is not restricted to dynamos, but it can also deal with decay problems,
as was already mentioned in Blackman \& Brandenburg (2002).
The dynamical quenching model is usually formulated in terms of $\alpha$,
but for helical mean fields $\meanJJ$ and $\meanBB$ are parallel and
the separation into contributions from $\alpha\meanBB$ and
$\eta_{\rm t}\meanJJ$ becomes less meaningful.
It is for this reasons that an $\alpha$ term appears in the description
of the decay of helical fields, rather than a dynamical contribution
to $\eta_{\rm t}$-quenching.

The remaining quenching of $\eta_{\rm t}$ that affects both helical
and nonhelical fields is consistent with an
algebraic quenching formula that is non-catastrophic, i.e.\ independent
of the microscopic magnetic diffusivity.

Although our results suggest that the turbulent magnetic Prandtl number
is of order unity, we cannot claim that it is safe to use turbulent
viscosity and magnetic diffusivity in a simulation of the dynamo or the
magnetorotational instability, for example, as a replacement of a fully
resolved simulation.
First of all, the functional form of the turbulent transport coefficients
is for realistic turbulent flows more complicated and involves in practice
tensorial rather than scalar coefficients.
Numerical evidence for this has been presented elsewhere in the context
of shear flow turbulence (Brandenburg \& Sokoloff 2002).
Furthermore, there will be additional terms such as the $\alpha$-effect
(see \Sec{Sdynquench}) and the AKA-effect (Frisch et al.\ 1987; see also
Brandenburg \& Rekowski 2001).
Most importantly, turbulent transport may be nonlocal, as is well known in
meteorology when modeling atmospheric flows (Stull 1984, Ebert et al.\ 1989),
where the turbulent transport is described by so-called transilient matrices
(see also Miesch et al.\ 2000 for examples of astrophysical convection).
Nonlocal transport means that the transport coefficients have to be
replaced by integral kernels.
In Fourier space, the convolution with an integral kernel corresponds
to a multiplication with a wavenumber dependent factor.
There is indeed some evidence that the main contribution comes only
from the smallest wavenumbers (Brandenburg \& Sokoloff 2002).
This is primarily a consequence of a lack of scale separation
in naturally forced turbulence, such as shear flows or convection.
In the present context, however, this is not an issue because we have
deliberately considered the case where the scale of the turbulent
eddies is much smaller than the scale of the large scale field
($k_{\rm f}/k_1=10$).

Finally, we wish to point out that studies of instabilities (e.g.\ the
magnetorotational or the dynamo instability) using turbulent transport
coefficients can sometimes lead to paradoxical situations.
In the case of solar convection, for example, one expects from mixing
length theory that turbulent viscosity and thermal diffusivity are on
the order of a few times $10^{12}\cm^2\s^{-1}$.
However, using such values in a global model of the sun leads to an
instability (R\"udiger 1989, R\"udiger \& Spahn 1992), which is in fact
nothing but a repetition of the original convection instability that leads
to turbulence in the first place (Tuominen et al.\ 1994).
It is therefore plausible that the actual values of the turbulent
transport coefficients should rather be close to the those for marginal
stability.
This would lead to a global constraint similar to the magnetic helicity
constraint that governs the nonlinear behavior of the $\alpha$-effect
in helical hydromagnetic turbulence.
At present, however, there is no theoretical framework that allows
self-consistent modeling of convection using mean-field theory.

\begin{acknowledgements}
We thank an anonymous referee for making useful suggestions and
drawing our attention to the paper by Fournier et al.\ (1982).
Use of the supercomputers in Odense (Horseshoe), Trondheim (Gridur),
and Leicester (Ukaff) is acknowledged.
\end{acknowledgements}


\vfill\bigskip\noindent\tiny\begin{verbatim}
$Header: /home/brandenb/CVS/tex/mhd/turb_prandtl/paper.tex,v 1.51 2004/02/02 17:49:53 tarek Exp $
\end{verbatim}

\end{document}